\documentclass[preprint,12pt]{elsarticle}

\usepackage{amssymb}
\usepackage{amstext}
\usepackage{graphicx}

\journal{Computer Physics Communications}

\begin{document}

\begin{frontmatter}



\title{On FFT--based convolutions and correlations, with application to solving Poisson's equation in an open rectangular pipe}


\author{Robert D. Ryne}

\address{Lawrence Berkeley National Laboratory,\\ 1 Cyclotron Road, Berkeley, CA 94720, United States}

\begin{abstract}
A new method is presented for solving Poisson's equation inside an open-ended rectangular pipe. The method uses Fast Fourier Transforms (FFTs) to perform mixed convolutions and correlations of the charge density with the Green function. Descriptions are provided for algorithms based on the ordinary Green function and for an integrated Green function (IGF). Due to its similarity to the widely used Hockney algorithm for solving Poisson's equation in free space, this capability can be easily implemented in many existing particle-in-cell beam dynamics codes.

\end{abstract}


\end{frontmatter}


\section{Introduction}
The solution of Poisson's equation is an essential component of any self-consistent beam dynamics code that models the transport of intense charged particle beams in accelerators,
as well as other plasma particle-in-cell (PIC) codes. If the bunch is small compared to the transverse size of the beam pipe, the conducting walls are usually neglected.
In such cases the Hockney method may be employed \cite{hockney, eastwoodandbrownrigg,hockneyandeastwood}. In that method, rather than computing $N_p^2$ point-to-point interactions
(where $N_p$ is the number of macroparticles), the potential is instead calculated on a grid of size $(2 N)^d$, where $N$ is the number of grid points in each dimension of the physical mesh containing
the charge, and where $d$ is the dimension of the problem.
Using the Hockney method, the calculation is performed using Fast Fourier Transform (FFT) techniques, with the computational effort scaling as $(2N)^d (log_2 2N)^d$.

When the beam bunch fills a substantial portion of the beam pipe transversely, or when the bunch length is long compared with the pipe transverse size, the conducting boundaries cannot be ignored. Poisson solvers have been developed previously to treat a bunch of charge in an open-ended pipe with various geometries \cite{qiangandryne,qiangandgluckstern}. Another approach (employed, {\it e.g.,} in the Warp code \cite{warp}), is to use a Poisson solver with periodic, Dirichlet, or Neumann boundary conditions on the pipe ends, and to extend the pipe in the simulation to be long enough so that the field is essentially zero there.

Here a new algorithm is presented for the open-ended rectangular pipe. The new algorithm is useful for a number of reasons. First, since its structure is very similar to the FFT-based free space method, it is straightforward to add this capability to any beam dynamics code that already contains the free space solver. Second, since it is Green-function based, the method does not require modeling the entire transverse pipe cross section, {\it i.e.}, if the beam was of small transverse extent one could instead model only a small transverse region around the axis.
Third, since it is based on convolutions and correlations involving Green functions, the method can use integrated Green function (IGF) techniques.
These techniques have the potential for higher efficiency and/or accuracy than non-IGF methods \cite{oxford,abelletal},
and are used in several beam dynamics codes including IMPACT, MaryLie/IMPACT, OPAL, ASTRA, and BeamBeam3D \cite{prstabquasistaticmodel,marylieimpact,opal,astra,beambeamcrab}.

The solution of the Poisson equation, $\nabla^2\phi=-\rho/\epsilon_0$, for the scalar potential, $\phi$, due to a charge density, $\rho$, can be expressed as,

\begin{equation}
\phi(x,y,z)=\int\int\int{dx' dy' dz'}\rho(x',y',z') G(x,x',y,y',z,z'),
\end{equation}
where $G(x,x',y,y',z,z')$ is the Green function, subject to the appropriate boundary conditions, describing the contribution of a source charge at location $(x',y',z')$ to the potential at an observation location $(x,y,z)$.
For an isolated distribution of charge this reduces to

\begin{equation}
\phi(x,y,z)=\int\int\int{dx' dy' dz'}\rho(x',y',z') G(x-x',y-y',z-z'),
\label{convolutionsolution}
\end{equation}
where

\begin{equation}
G(u,v,w)={1\over \sqrt{u^2+v^2+w^2}}.
\label{isolatedgreenfunction}
\end{equation}
A simple discretization of Eq.~(\ref{convolutionsolution})
on a Cartesian grid with cell size $(h_x,h_y,h_z)$
leads to,

\begin{equation}
\phi_{i,j,k}=h_x h_y h_z \sum_{i'=1}^{i'_{max}}\sum_{j'=1}^{j'_{max}}\sum_{k'=1}^{k'_{max}}  \rho_{i',j',k'}G_{i-i',j-j',k-k'},
\label{openbruteforceconvolution}
\end{equation}
where $\rho_{i,j,k}$ and $G_{i-i',j-j',k-k'}$ denote the values of the charge density and the Green function, respectively, defined on the grid.

As is well known \cite{numericalrecipes}, FFT's can be used to compute convolutions by appropriate zero-padding of the sequences.
A proof of this is shown in the Appendix, along with an explanation of the requirements for the zero-padding.
As a result, the solution of Eq.~(\ref{openbruteforceconvolution}) is then given by

\begin{equation}
\phi_{i,j,k}=h_x h_y h_z \mathcal{F}^{bbb} \{ (\mathcal{F}^{fff}\rho_{i,j,k}) (\mathcal{F}^{fff}G_{i,j,k}) \}
\label{oneterm}
\end{equation}
where the notation has been introduced that $\mathcal{F}^{fff}$ denotes a forward FFT in all 3 dimensions,
and $\mathcal{F}^{bbb}$  denotes a backward FFT in all 3 dimensions.
The treatment of the open rectangular pipe relies on the fact, as described in the Appendix, that the FFT-based approach
works for correlations as well as for convolutions, the only difference being the direction of the FFTs.
As an example, consider a case for which the $y$ variable involves a correlation instead of a convolution. Then,

\begin{equation}
\sum_{i'=1}^{i'_{max}}\sum_{j'=1}^{j'_{max}}\sum_{k'=1}^{k'_{max}} \rho_{i',j',k'}G_{i-i',j+j',k-k'}=\mathcal{F}^{bfb} \{ (\mathcal{F}^{fff}\rho_{i,j,k}) (\mathcal{F}^{fbf}G_{i,j,k}) \}.
\label{mixed}
\end{equation}
Because of this, and the fact the the Green function for a point charge in an open rectangular
pipe is a function of $(x\pm x', y \pm y', z\pm z')$, an FFT-based algorithm follows immediately.

\section{Poisson's equation in an open rectangular pipe}
The Green function for a point charge in an open rectangular pipe with transverse size $(0,a)\times(0,b)$
can be found using eigenfunction expansions (see, {\it e.g.,} \cite{jackson}). It can be expressed as,

\begin{equation}
G(x,x',y,y',z,z')={1\over 2\pi a b}\sum_{m=1}^{\infty}\sum_{n=1}^{\infty}{1\over \kappa_{mn}}\sin{{m \pi x \over a}}\sin{{m \pi x' \over a}}\sin{{n \pi y \over b}}\sin{{n \pi y' \over b}} e^{-\kappa_{mn}|z-z'|},
\end{equation}
where $\kappa_{mn}^2=({m\pi\over a})^2+({n\pi\over b})^2$ ~\cite{smythe}.
Equivalently, it can be expressed as a function of a single rectangular pipe Green function,
\begin{eqnarray}
G&=&~~R(x-x',y-y',z-z') - R(x-x',y+y',z-z') \nonumber \\
   &&-R(x+x',y-y',z-z') + R(x+x',y+y',z-z'),
\label{green4terms}
\end{eqnarray}
where
\begin{equation}
R(u,v,w)={1\over 2\pi a b}\sum_{m=1}^{\infty}\sum_{n=1}^{\infty}{1\over \kappa_{mn}}\cos{{m \pi u \over a}}\cos{{n \pi v \over b}}e^{-\kappa_{mn}|w|}.
\label{rgreenfunction}
\end{equation}
It follows that an algorithm for solving the Poisson equation in an open rectangular pipe is given by
\begin{eqnarray}
&&\phi_{i,j,k}/(h_x h_y h_z )= \nonumber \\
&&~~\mathcal{F}^{bbb} \{ (\mathcal{F}^{fff}\rho_{i,j,k}) (\mathcal{F}^{fff}R_{i,j,k})\} -
     \mathcal{F}^{bfb} \{ (\mathcal{F}^{fff}\rho_{i,j,k}) (\mathcal{F}^{fbf}R_{i,j,k})\}  \nonumber \\
&&-\mathcal{F}^{fbb} \{ (\mathcal{F}^{fff}\rho_{i,j,k}) (\mathcal{F}^{bff}R_{i,j,k})\} +
    \mathcal{F}^{ffb} \{ (\mathcal{F}^{fff}\rho_{i,j,k}) (\mathcal{F}^{bbf}R_{i,j,k})\}
\label{fourterms}
\end{eqnarray}
At each step of a simulation
charge is deposited on a doubled grid, and its forward FFT, $(\mathcal{F}^{fff}\rho_{i,j,k})$, is computed.
The function $R$ is tabulated in the doubled domain, and 4 mixed Fourier transforms
are computed, namely $(\mathcal{F}^{fff}R_{i,j,k})$, $(\mathcal{F}^{fbf}R_{i,j,k})$, $(\mathcal{F}^{bff}R_{i,j,k})$,
and $(\mathcal{F}^{bbf}R_{i,j,k})$.
The transformed charge density is multiplied by each of
the transformed Green functions, 4 final FFTs are performed, and the results are added
to obtain the potential.

Note that, although the Green function in Eq.~(\ref{green4terms}) contains 4 terms, only a {\em single} function $R$ needs to be tabulated on a grid,
and it is this tabulated function that is transformed in 4 different ways and stored.

\section{Integrated Green Function}
The discrete convolution, Eq.~(\ref{openbruteforceconvolution}), is a simple approximation to Eq.~(\ref{convolutionsolution}).
It makes use of the Green function only at the grid points, even though it is known everywhere in space.
This can lead to serious inaccuracy when $\rho$ and $G$ have a disparate spatial variation,
especially when the
grid cells have high aspect ratio.
This can happen in a variety of situations (isolated, open-ended pipe, {\it etc.}) depending on the problem geometry and the number of grid cells in each dimension.
For example, a simulation of a disk-shaped galaxy would have highly elongated cells unless many more cells were used in the wide direction.
Also, inside a conducting pipe, the beam might be long
and slowly varying, whereas the pipe Green function decays exponentially with $z$.

Qiang described a method for approximating Eq.~(\ref{convolutionsolution}) to arbitrary accuracy using the Newton-Cotes formula \cite{qiangnewtoncotes}.
Integrated Green functions (IGF's) provide another means to approximate Eq.~(\ref{convolutionsolution}) accurately when certain integrals involving the Green function can be obtained analytically \cite{oxford,abelletal}.
This is accomplished by assuming a simple analytical form for the variation of $\rho$ within a cell, and analytically performing the convolution (or correlation) integrals for each cell of the problem.
As a result, the accuracy is controlled by how well the discretization resolves $\rho$, not $G$.

For illustration, consider the 2D problem with isolated boundary conditions. If $\rho$ were assumed to be a sum of delta functions at the grid points, we would obtain the 2D analog of Eq.~(\ref{openbruteforceconvolution}).
Instead suppose that linear basis functions are used to approximate $\rho$ within each cell.
Then
\begin{eqnarray}
&&\phi(x_i,y_j) = \nonumber \\
&~~&\sum_{i',j'}  \rho_{i',j'} \int_{0}^{h_x}\int_{0}^{h_y}dx'dy'~{(h_x-x')(h_y-y')\over h_x h_y} G(x_{i}-x_{i'}-x',y_{j}-y_{j'}-y') \nonumber \\
&+ &\sum_{i',j'}  \rho_{i'+1,j'}\int_{0}^{h_x}\int_{0}^{h_y}dx'dy'~{x'(h_y-y')\over h_x h_y} G(x_{i}-x_{i'}-x',y_{j}-y_{j'}-y') \nonumber \\
&+ &\sum_{i',j'}  \rho_{i',j'+1}\int_{0}^{h_x}\int_{0}^{h_y}dx'dy'~{(h_x-x')y'\over h_x h_y} G(x_{i}-x_{i'}-x',y_{j}-y_{j'}-y') \nonumber \\
& +& \sum_{i',j'}   \rho_{i'+1,j'+1}\int_{0}^{h_x}\int_{0}^{h_y}dx'dy'~{x' y'\over h_x h_y} G(x_{i}-x_{i'}-x',y_{j}-y_{j'}-y').
\label{igf2dintegrals}
\end{eqnarray}
Shifting the indices in the last 3 sums above, collecting terms, and factoring out $h_x h_y$ (in analogy with Eq.~(\ref{openbruteforceconvolution})),
the IGF is the coefficient of $h_x h_y \rho_{i',j'}$~,
\begin{equation}
\phi(x_i,y_j) =h_x h_y \sum_{i',j'}  \rho_{i',j'} G^{int}_{i-i',j-j'}.
\label{igf2dconvolution}
\end{equation}
To simplify the discussion that follows, suppose that $\rho$ is zero on the boundary; this allows one to shift indices and collect terms without
worrying about possible boundary issues.

Returning to the problem of the open-ended rectangular pipe, here we will use the IGF approach to treat the longitudinal dimension.
Suppose that, inside the $k^{th}$ cell,
the longitudinal dependence of the charge density is given by,
\begin{equation}
\rho(z)= {1 \over h_z} \left[\rho_{k}(h_z-(z-z_k))+\rho_{k+1}~(z-z_k)\right].
\label{lineardensity}
\end{equation}
The longitudinal dependence of the IGF is therefore,
\begin{equation}
g_z= {1 \over h_z^2}\int_{z_k}^{z_{k+1}}dz' \left[\rho_{k}(h_z-(z'-z_k))+\rho_{k+1}~(z'-z_k)\right]e^{-\kappa_{mn}|z-z'|}.
\label{twoterms}
\end{equation}
Performing the integrations and summing over all the cells, two terms from adjacent cells contribute to the the IGF,
with the result,
\begin{equation}
g_z(w)={1\over h_z^2 \kappa_{mn}^2} \Big[  2 h_z \kappa_{mn} \delta_{w,0} + \left(e^{-\kappa_{mn}|w+h_z|}-2e^{-\kappa_{mn}|w|}+e^{-\kappa_{mn}|w-h_z|}\right)\Big],
\label{gsubz}
\end{equation}
where $w=z-z_k$.
In summary, in analogy to Eq.~(\ref{rgreenfunction}), the integrated Green function, $R_{int}$, integrated in just the longitudinal coordinate, for a distribution of charge in an open-ended rectangular pipe is given by,
\begin{equation}
R_{int}(u,v,w)={1\over 2\pi a b}\sum_{m=1}^{\infty}\sum_{n=1}^{\infty}{1\over \kappa_{mn}}\cos{{m \pi u \over a}}\cos{{n \pi v \over b}}g_z(w).
\label{rintgreenfunction}
\end{equation}

\section{Numerical Example}
\begin{figure}[h]
\begin{tabular}{cc}
\includegraphics[height=1.8in]{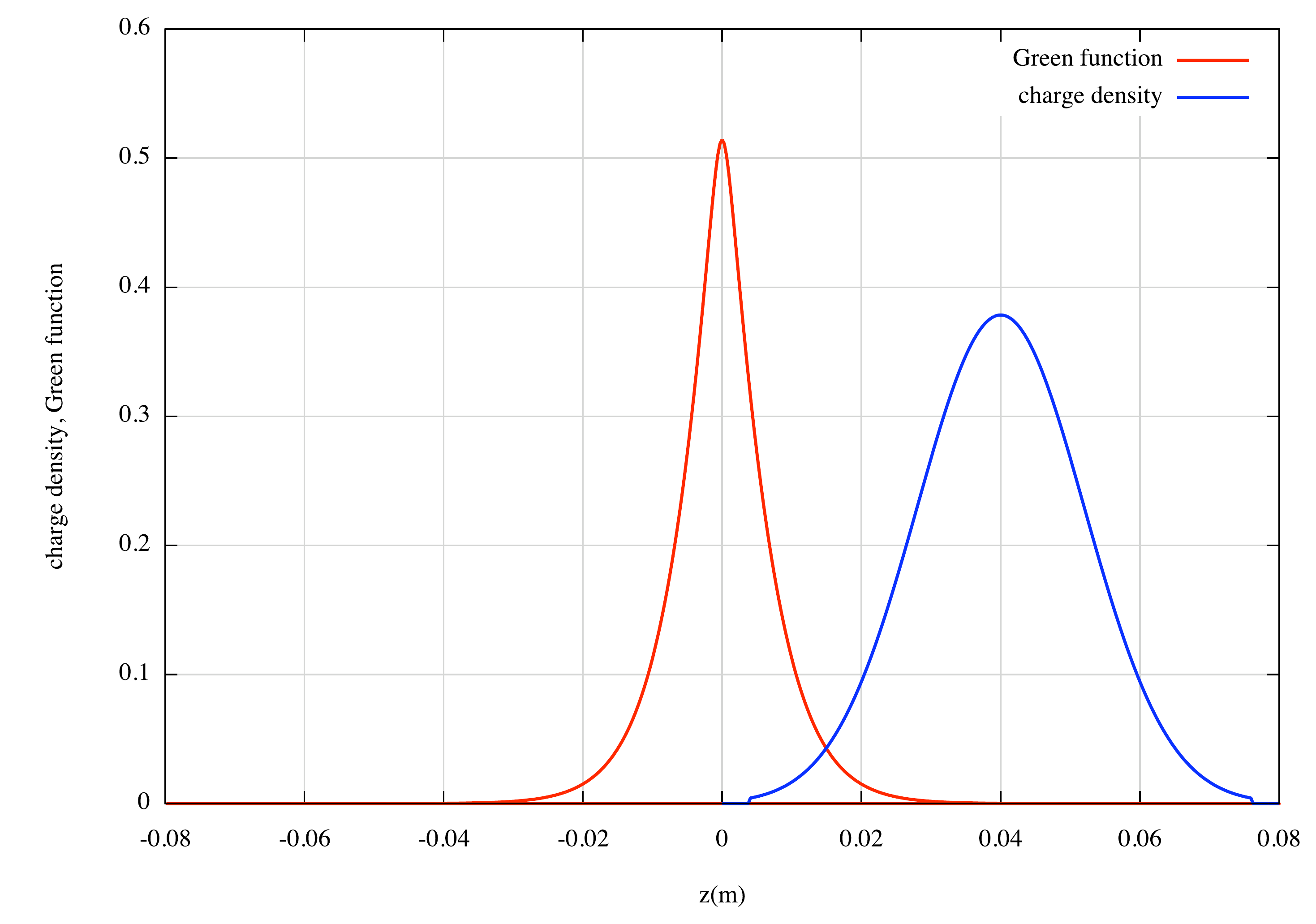} &
\includegraphics[height=1.8in]{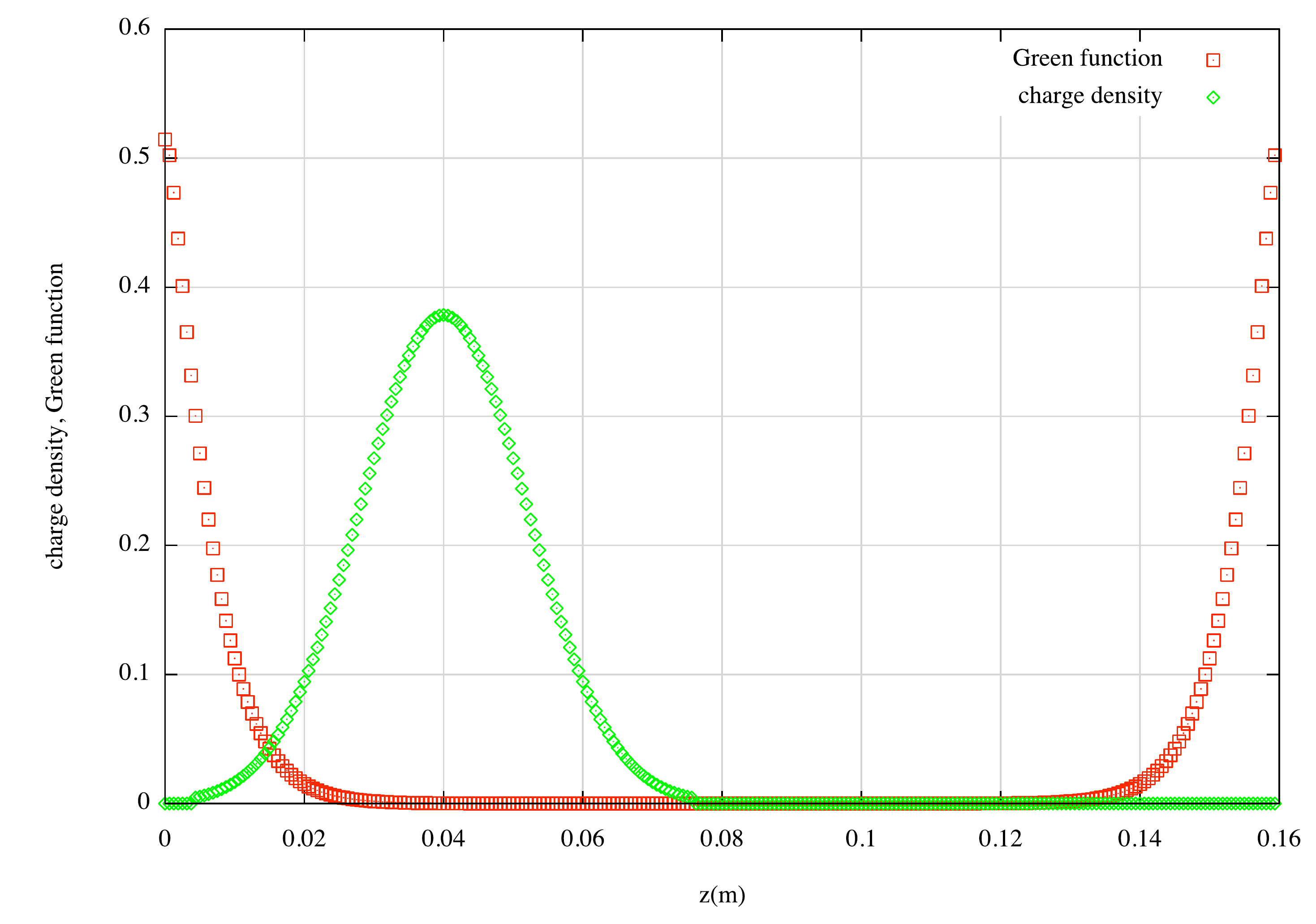} \\
\end{tabular}
\caption{Left: On-axis charge density and Green function vs. z, as they appear in physical space, for the case $\sigma_z=1.2~\rm cm$.
                Right: On-axis charge density and Green function as stored in memory.
                }\label{fig1label}
\end{figure}

Consider a rectangular waveguide of full width and height $a=b=4~\rm cm$.
The Green functions, Eq.~(\ref{rgreenfunction}) and Eq.~(\ref{rintgreenfunction}), will be calculated using $m,n=1,\ldots,20$.
Consider a 3D Gaussian charge distribution with transverse rms sizes $\sigma_x=0.15a=6\rm mm$, $\sigma_y=0.15b=6\rm mm$, and longitudinal rms size $\sigma_z$.
Three cases with different rms bunch length will be considered,  $\sigma_z=1.2~\rm cm$, $\sigma_z=12~\rm cm$, and $\sigma_z=1.2~\rm m$.
The distribution is set to zero at $x^2/\sigma_x^2+y^2/\sigma_y^2+z^2/\sigma_z^2 > 3^2$.
Fig.~\ref{fig1label}, left, shows the charge density and Green function as a function of $z$, down the center of the pipe, for the case $\sigma_z=1.2~\rm cm$.
As shown in the Appendix, the convolution can be performed using FFTs by zero-padding the charge density over the domain of the Green function, and by treating the Green function as a periodic function.

Fig.~\ref{fig1label}, right, shows how the charge density and Green function are actually stored in memory,
using 256 longitudinal grid points: $\rho$ is left in place and zero-padded, and $G$ is circular-shifted so that $G(x=0,y=0,z=0)$ is at the lower bound of the Green function array.
For the case $\sigma_z=12~\rm cm$, Fig.~\ref{fig1label}, right, would look similar except that the Green function would be confined to narrow regions at the left and right edges of the plot.
For the case $\sigma_z=1.2~\rm m$, it would be confined to extremely narrow regions at the left and right edges.
 
\begin{figure}[h]
\begin{tabular}{cc}
\includegraphics[height=1.8in]{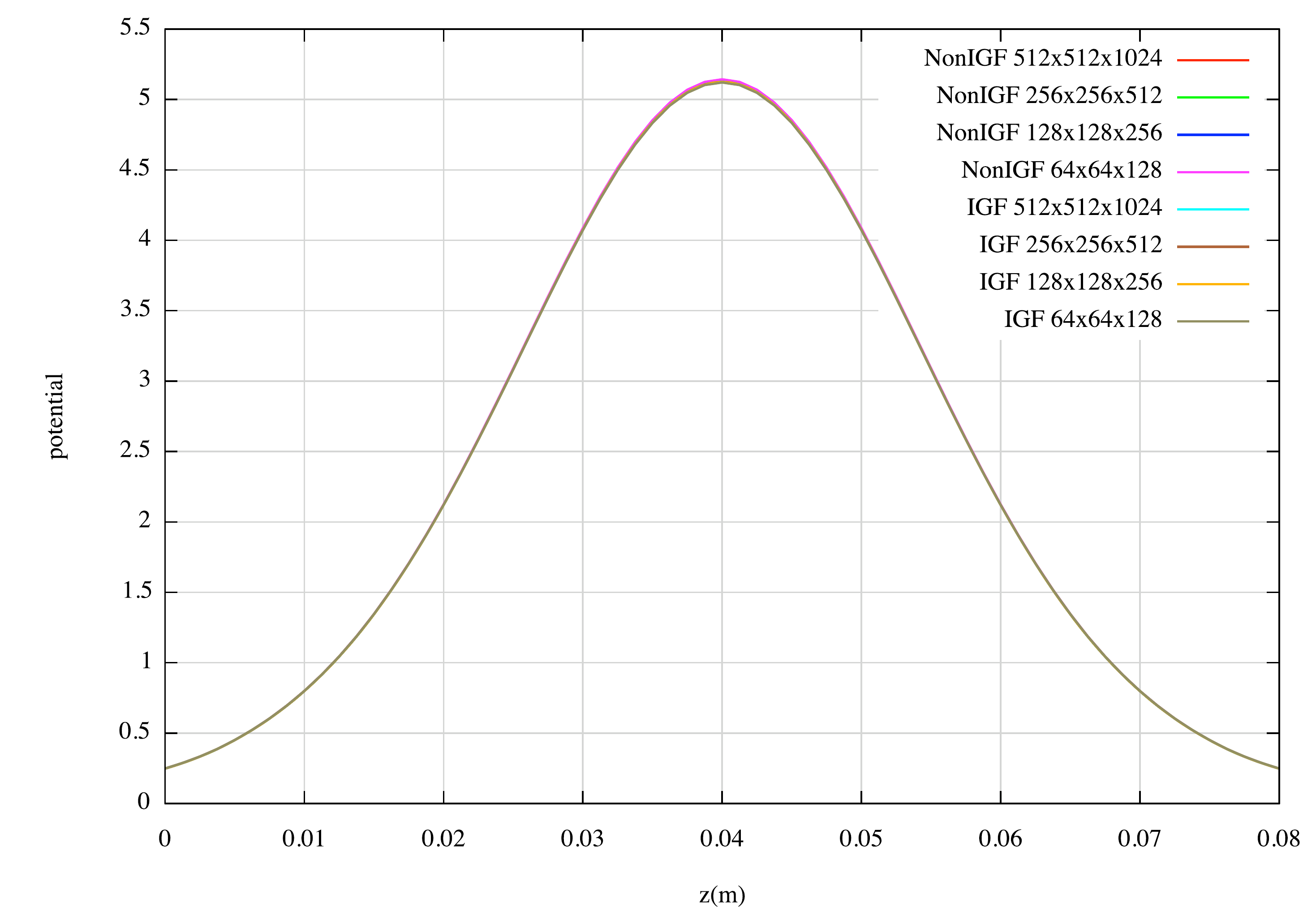} &
\includegraphics[height=1.8in]{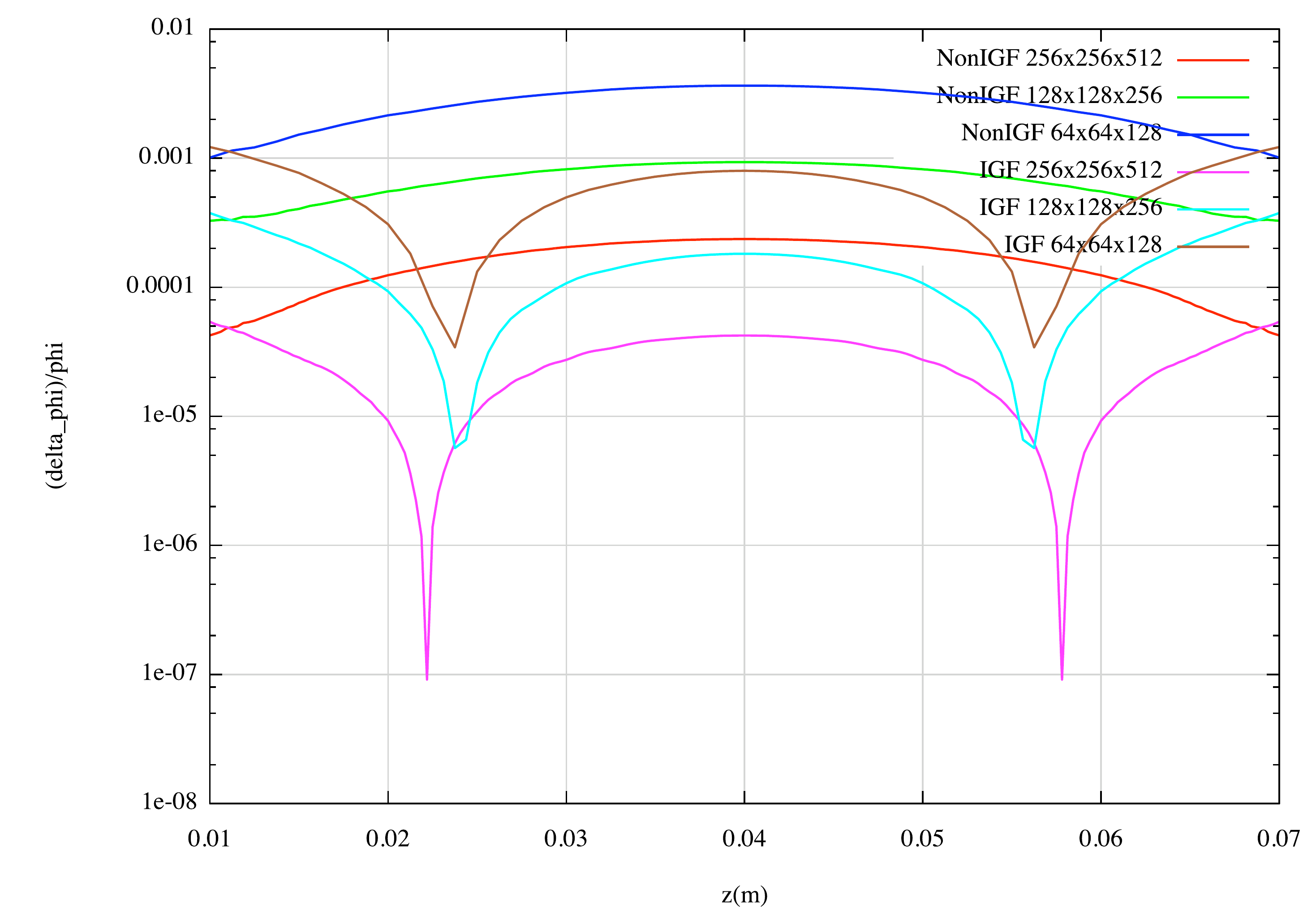} \\
\end{tabular}
\caption{Left: On-axis potential vs. z showing the ordinary Green function result and the Integrated Green function (IGF) result for various grid sizes.
 The bunch is a Gaussian distribution with $\sigma_x=\sigma_y=6\rm mm$, $\sigma_z=1.2~\rm cm$.
                Right: Relative error of the on-axis potential vs. z for grid sizes $64\times 64 \times 128$, $128\times 128 \times 256$, and $256\times 256 \times 512$.
                }\label{fig2label}
\end{figure}

\begin{figure}[h]
\begin{tabular}{cc}
\includegraphics[height=1.8in]{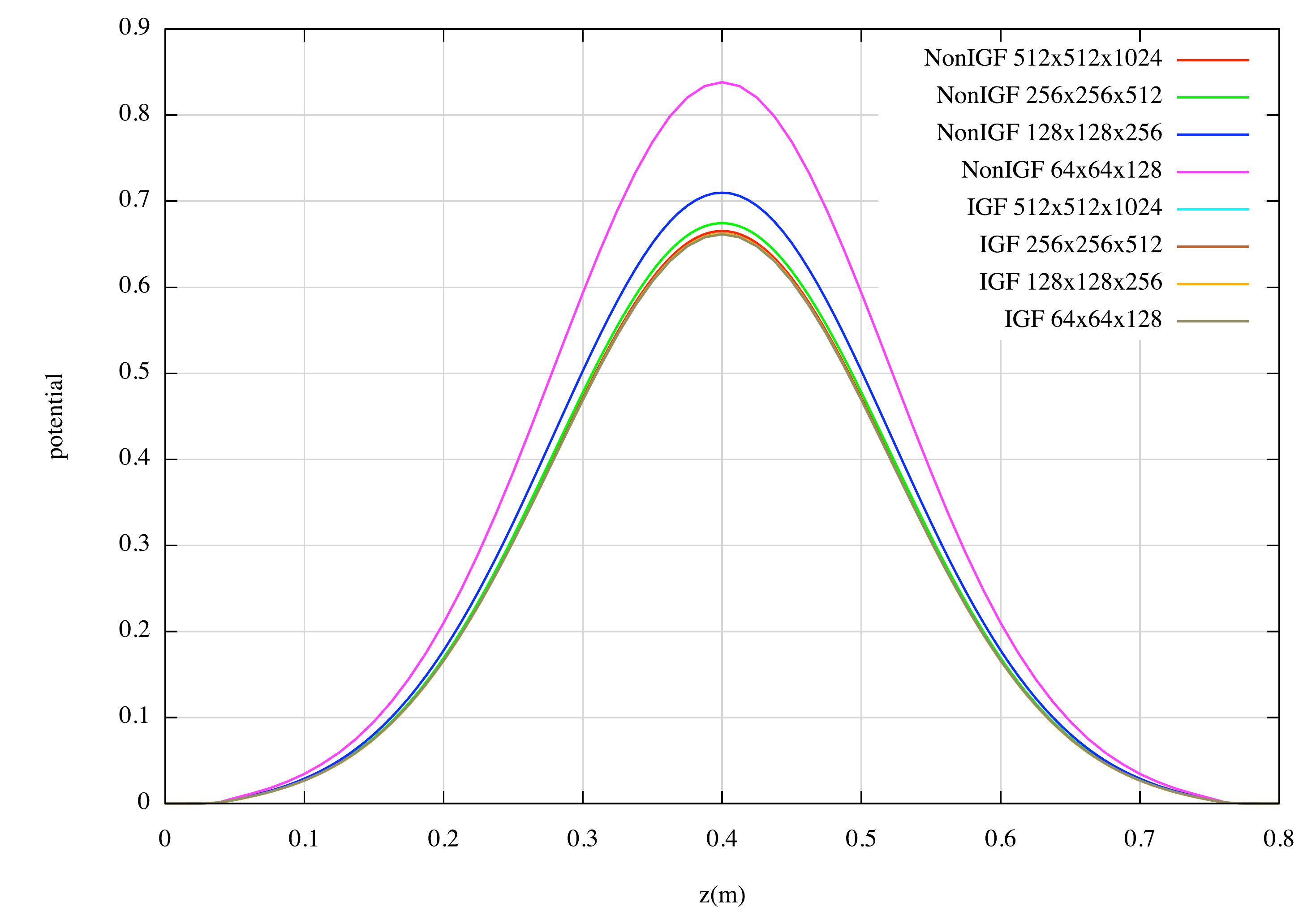} &
\includegraphics[height=1.8in]{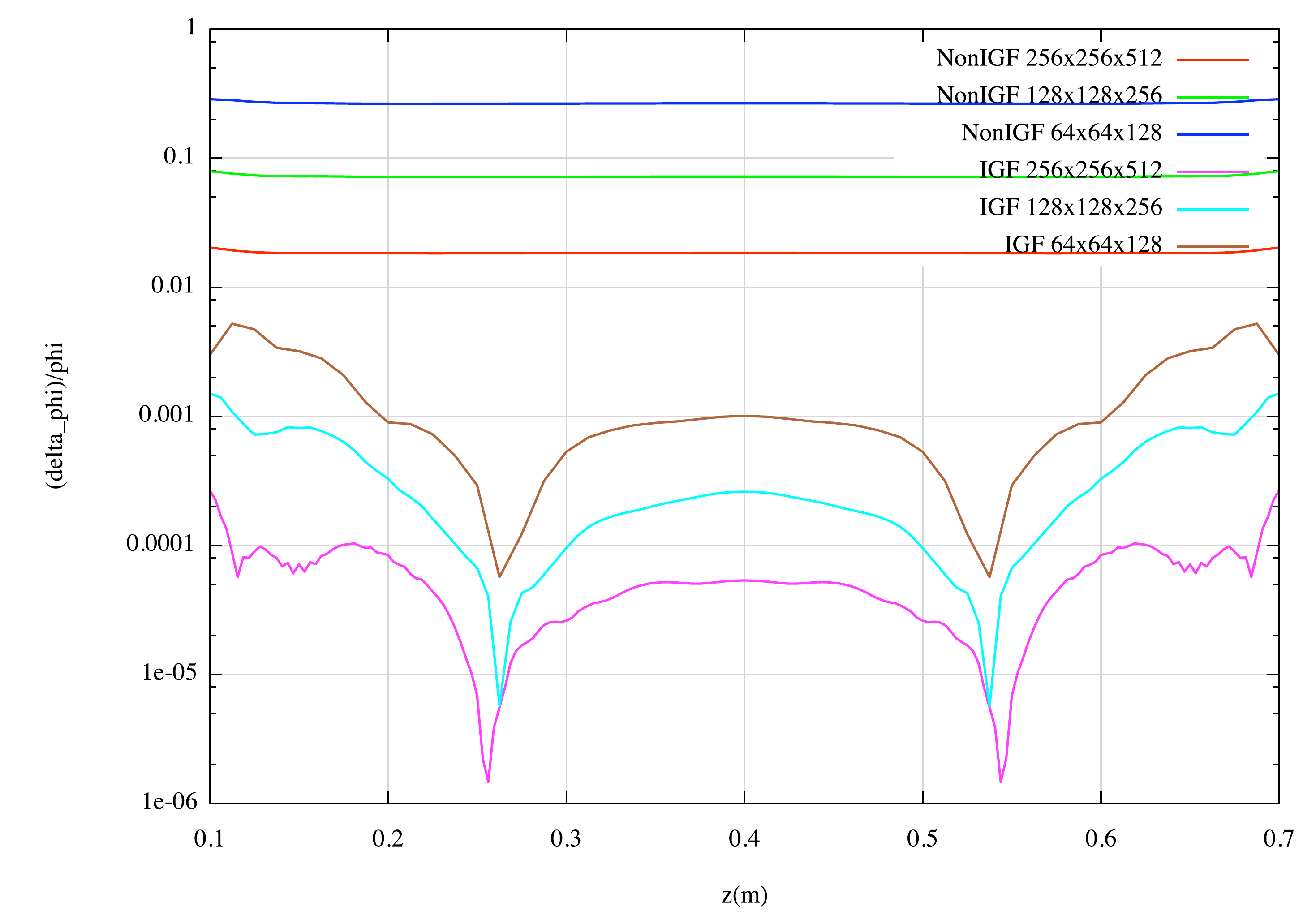} \\
\end{tabular}
\caption{Left: On-axis potential vs. z showing the ordinary Green function result and the IGF result for various grid sizes.
 The bunch is a Gaussian distribution with $\sigma_x=\sigma_y=6\rm mm$, $\sigma_z=12~\rm cm$.
                Right: Relative error of the on-axis potential vs. z.
                }\label{fig3label}
\end{figure}

\begin{figure}[h]
\begin{tabular}{cc}
\includegraphics[height=1.8in]{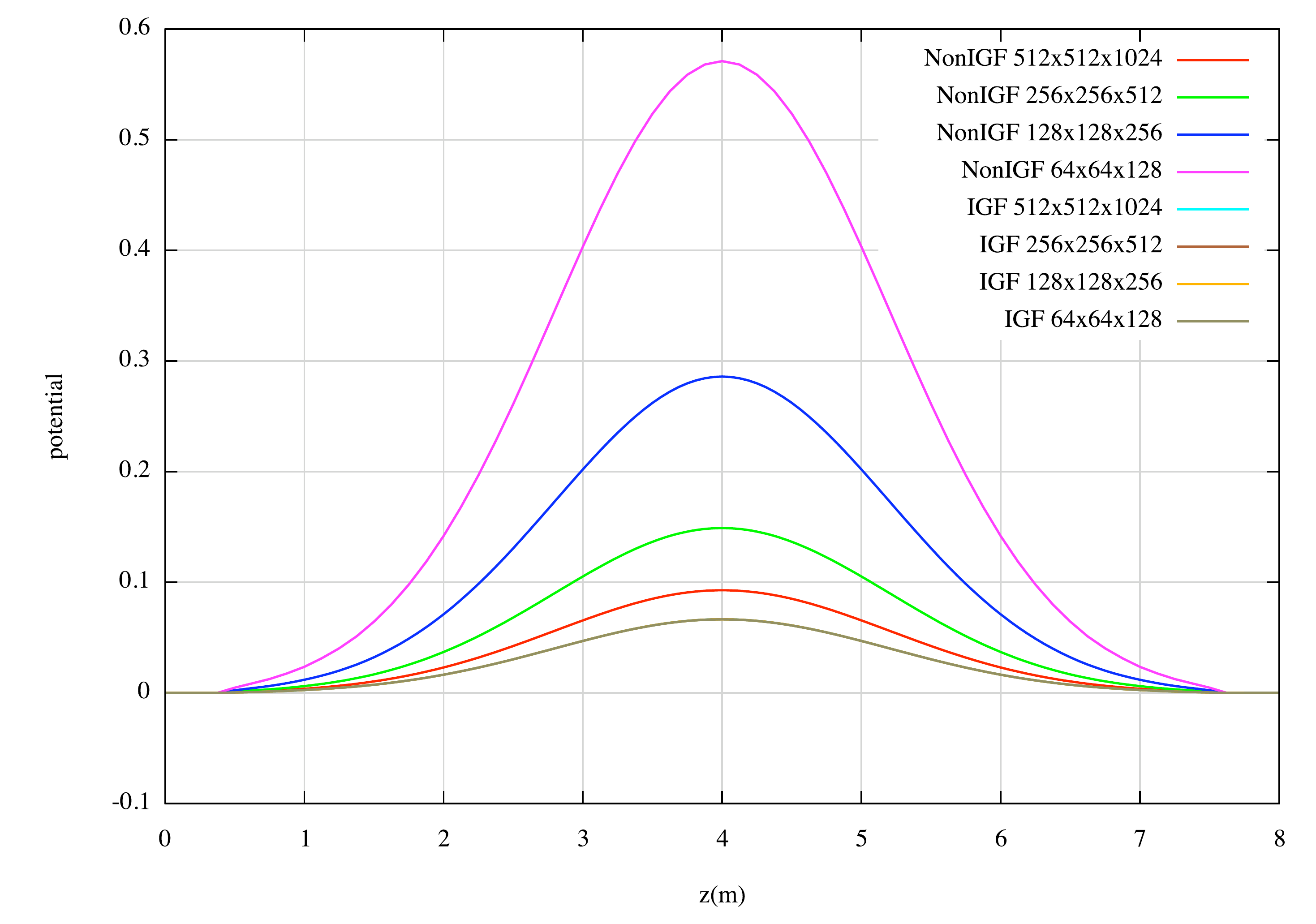} &
\includegraphics[height=1.8in]{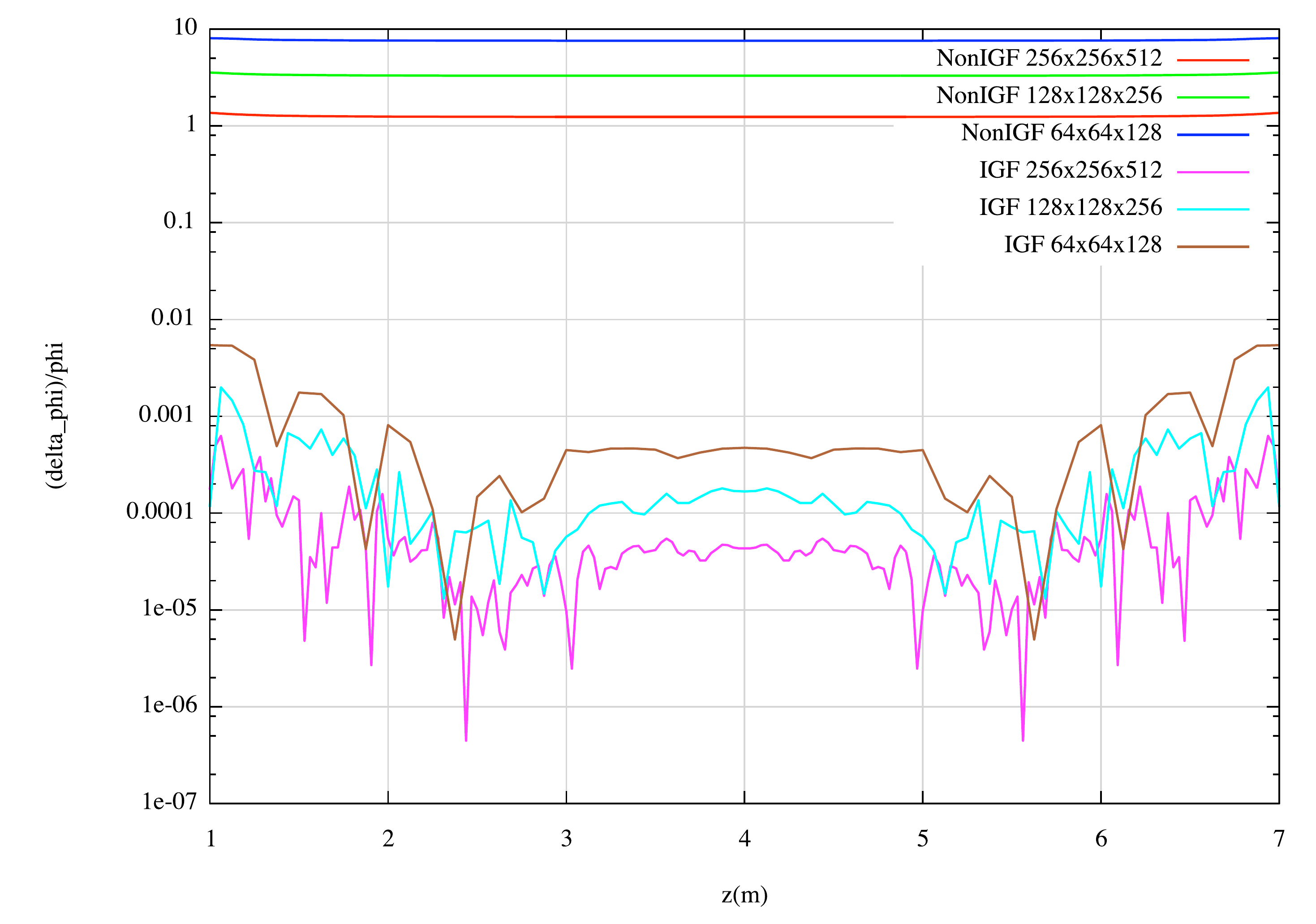} \\
\end{tabular}
\caption{Left: On-axis potential vs. z showing the ordinary Green function result and the IGF result for various grid sizes.
 The bunch is a Gaussian distribution with $\sigma_x=\sigma_y=6\rm mm$, $\sigma_z=1.2~\rm m$.
                Right: Relative error of the on-axis potential vs. z.
                }\label{fig4label}
\end{figure}

Figures 2, 3, and 4 show convolution results for the three different bunch lengths,  $\sigma_z=1.2~\rm cm$, $\sigma_z=12~\rm cm$, and $\sigma_z=1.2~\rm m$, respectively,
for grid sizes $64\times 64\times 128$ up to $512\times 512 \times 1024$.
The left hand side of each figure shows plots of the potential as a function of $z$ on-axis for various grid sizes, comparing results based on the ordinary Green function and the integrated Green function.
The right hand side shows the relative error of the calculated potential.
In Fig.~\ref{fig2label}, $\sigma_z$ is less than the pipe transverse size $(1.2~\rm{cm}~vs.~ 4~\rm{cm})$; both the ordinary Green function and the IGF are accurate to better than 1\% for all the grid sizes shown.
In Fig.~\ref{fig3label}, $\sigma_z$ is somewhat larger than the pipe transverse size $(12~\rm{cm}~vs.~ 4~\rm{cm})$;
when the grid is coarse, the ordinary Green function has significant errors (a few percent to several tens of percent), while the IGF accuracy is 1\% or better.
In Fig.~\ref{fig4label}, $\sigma_z$ is much larger than the pipe transverse size $(1.2~\rm{m}~vs.~ 4~\rm{cm})$;
in this case when the grid is coarse the ordinary Green function results exhibit huge errors (more than 100\%), while the IGF accuracy is still 1\% or better.
As mentioned above, the accuracy of the IGF results is controlled by how well the grid resolves {\em just} the charge density. 
For the non-IGF results, the accuracy depends on resolving {\em both} the charge density and Green function, and, due to the exponential fall-off of the Green function, a coarse
grid gives unusable results. The relative error in the potential is plotted on the right hand side of the figures. These were obtained be plotting $(\phi-\phi_{highres})/\phi_{highres})$,
where $\phi_{highres}$ is the highest resolution result, obtained using the IGF with a $512\times 512 \times 1024$.


\begin{figure}[h]
\begin{tabular}{cc}
\includegraphics[height=1.8in]{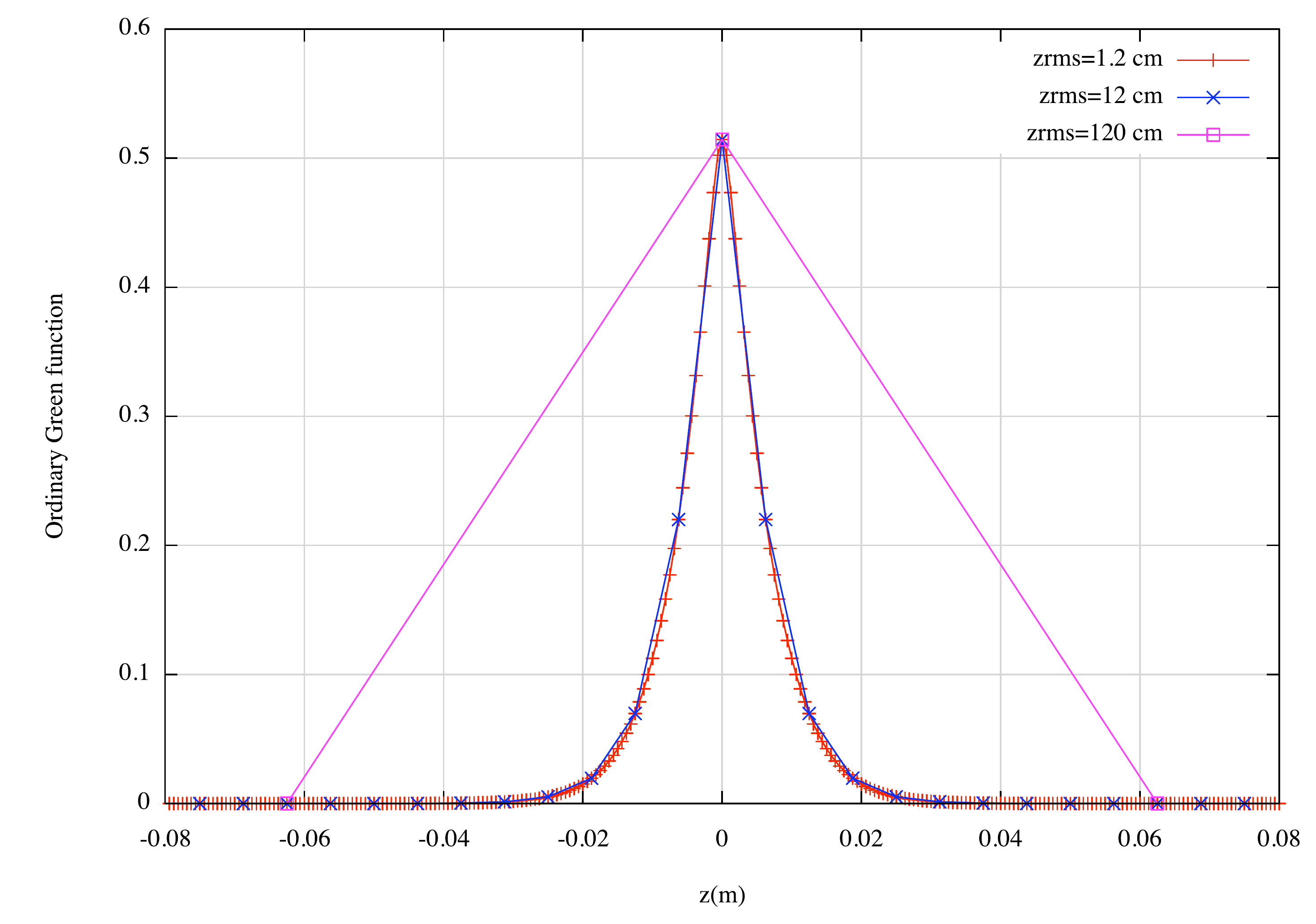} &
\includegraphics[height=1.8in]{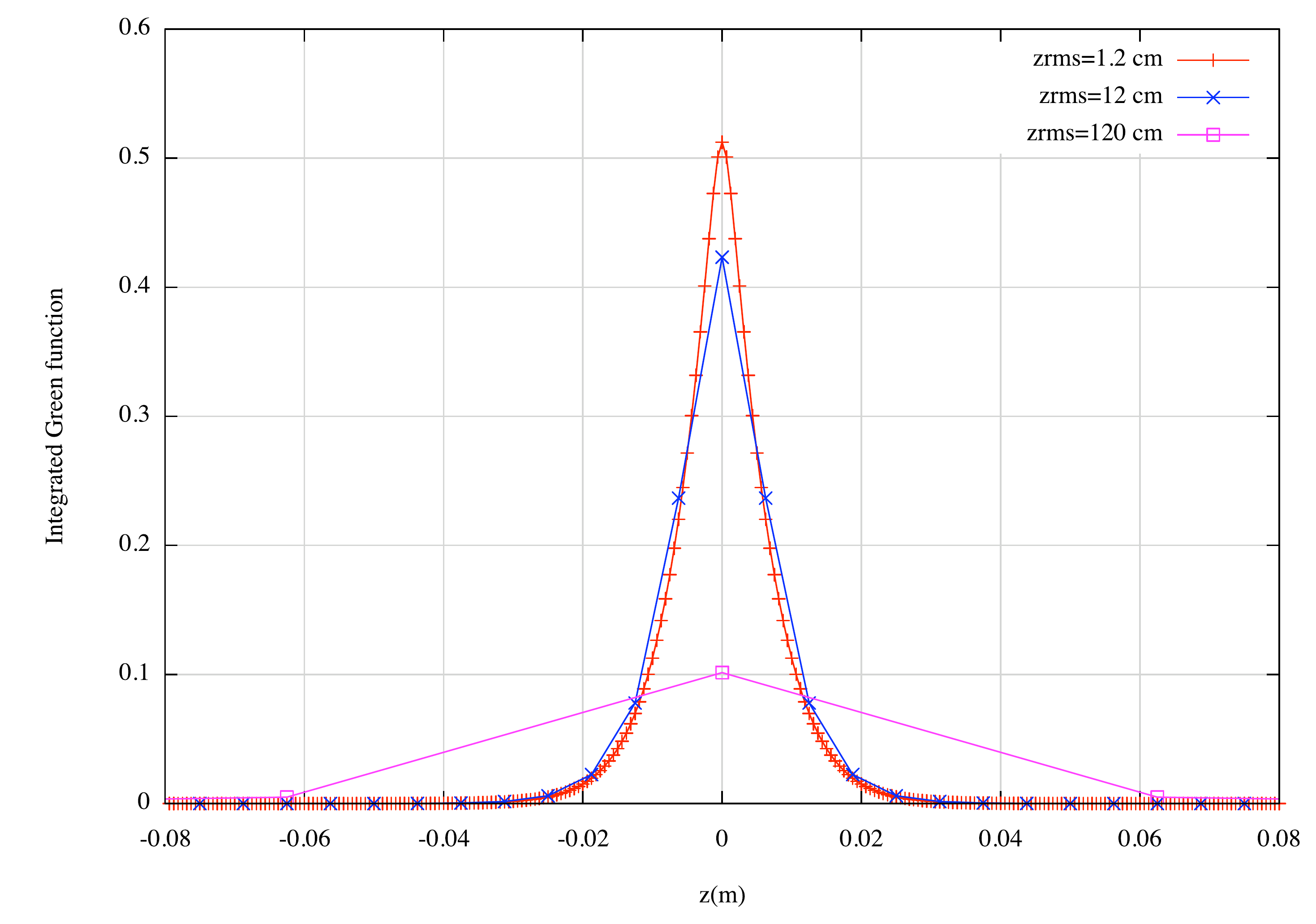} \\
\end{tabular}
\caption{Left: Ordinary Green function vs. z, for three different bunch lengths, using 256 grid points.
                Right: Integrated Green function vs. z.
                }\label{fig5label}
\end{figure}

Fig.~\ref{fig5label} shows the ordinary Green function and the IGF on-axis in the range $z\in[-8~cm,8~cm]$ for the three cases  $\sigma_z=1.2~\rm cm$,  $\sigma_z=12~\rm cm$, and  $\sigma_z=1.2~\rm m$,
all computed using 256 longitudinal grid points.
The left hand side shows the ordinary Green function. The exponential fall-off of the Green function is obvious.
Note that the ordinary Green function is the {\em same} for all 3 cases, it is just sampled differently.
Note also that for $\sigma_z=1.2~\rm m$, the grid is so coarse that only the central point is significant, all the others are so far into the region of exponential fall-off that they are effectively zero.
 If $\rho$ were constant inside of a cell (which is approximately true for $\sigma_z=1.2~\rm m$), use of Eq.~(\ref{openbruteforceconvolution}) would correspond to approximating
 the area under the true Green function with the area under the triangular shaped region of Fig.~\ref{fig5label}, left;
 so it is no surprise that the non-IGF result in Fig.~\ref{fig4label} is highly inaccurate, with errors exceeding $100\%$.
The right hand side of Fig.~\ref{fig5label} shows the IGF on-axis. As before, the exponential fall-off of the Green function is obvious.
 However, it does not matter that the coarse data poorly resolves the IGF, because the accuracy does not depend on that, it only depends on having a fine enough grid that
 the charge density is well approximated by a linear function within a cell. For $\sigma_z=1.2~\rm m$, this approximately true even with as few as 128 grid points,
 hence the IGF result in Fig.~\ref{fig4label} has good accuracy.
 
 Lastly note that, for $\sigma_z=1.2 \rm m$ with 256 grid points, 
 the IGF is approximately equal to zero for all but the central point. Hence,
 the four terms in Eq.~(\ref{fourterms}) for the discrete convolution/correlation solution of the Poisson equation, such as 
 the term on the left-hand side of Eq.~(\ref{mixed}), could be approximated by keeping just $k'=0$ in the sum over $k'$,
 \begin{equation}
h_x h_y h_z \sum_{i'=1}^{i'_{max}}\sum_{j'=1}^{j'_{max}} \rho_{i',j',k}G_{i-i',j+j',0}.
\label{openbruteforceconvolution2d}
\end{equation}
In other words, the problem has been converted to one involving multiple 2D -- instead of 3D -- mixed convolutions and correlations.
The would reduce execution time by a factor equal to $\rm log_2$ of the number of longitudinal grid points.
More generally, depending on the grid size and the exponential fall-off, it would be sufficient compute,
\begin{equation}
h_x h_y h_z \sum_{i'=1}^{i'_{max}}\sum_{j'=1}^{j'_{max}}\sum_{k'=k}^{k\pm N_{neighbors}}  \rho_{i',j',k'}G_{i-i',j+j',k-k'},
\label{neighboringconvolution}
\end{equation}
where $N_{neighbors}=0$ if the IGF, $G_{i,j,k}$~, is appreciable only when $k=0$, or $N_{neighbors}=1$ if the IGF is appreciable at  $k=0$ and $k=\pm 1$,
{\it etc}.
 
\section{Discussion and Conclusion}

A new method has been presented for solving Poisson's equation in an open-ended rectangular pipe.
Compared with the Hockney method for isolated systems (Eq.~(\ref{oneterm})) which can be computed with a single FFT-based convolution,
the new method involves 4 mixed convolutions and correlations (Eq.~(\ref{fourterms})), where the convolutions and correlations are mixed
among the problem dimensions. Starting with the Green function for a charge in an open-ended rectangular pipe (Eq.~(\ref{rgreenfunction})),
an Integrated Green function (IGF) was derived (Eq.~(\ref{rintgreenfunction})). Simulations of a Gaussian beam in an open-ended pipe
showed that the IGF approach is much more robust, {\it i.e.,} it retains much better accuracy, than the non-IGF method over a wide range of bunch lengths.
This is because the accuracy of the IGF approach depends only on having a fine enough grid resolve the spatial variation of the charge density.
In contrast, the non-IGF approach is sensitive to disparities between the spatial variation of the Green function and the charge density.
Since the effort to compute the IGF
is not much more than for the ordinary Green function,
this argues that one should always use the IGF.

In theory the calculation of the IGF, which is represented as an infinite series in Eq.~(\ref{rintgreenfunction}), could
make the simulation much more time consuming than the case for isolated boundary conditions.
But in practice this is unlikely since simulations with isolated boundary conditions often re-grid
at every time step (or as needed) to take account of the changing beam size.
This would not be the case for the rectangular pipe Poisson solver if the beam filled most of the pipe transversely,
since the IGF would be computed once, Fourier transformed 4 ways, stored, and reused.
A possible exception would be if the longitudinal size were changing rapidly with time, but that is not usually the case.
In fact, in many applications involving circular accelerators the longitudinal bunch oscillations are very slow.
It should also be pointed out that it is not always necessary to compute the Green function over the full transverse cross section.
If the beam happened to be very small transversely (but long compared with the transverse pipe dimensions so that
the open treatment would be inappropriate), then the transverse physical mesh could be a small
area centered on the z-axis,
which would be much more efficient and accurate than gridding the full cross section.

To implement this approach one needs to be able to control the direction of the FFTs in all dimensions.
Unfortunately most packages don't offer this capability for built-in multidimensional FFTs
(an exception was the Connection Machine Scientific Software Library),
so it is up to the programmer to implement it using 1D FFTs.
Such a capability has other uses. For example, when dealing with the Vlasov equation,
\begin{equation}
{\partial f \over \partial t} + ({\vec p}\cdot \partial_{\vec q})f -  ({\nabla\Phi}\cdot \partial_{\vec p})f=0,
\end{equation}
a split-operator, FFT-based method for advancing the distribution in phase space one time step can be expressed as \cite{rynehabibwangler},
\begin{equation}
f({\vec q},{\vec p},t)=e^{-{t\over 2}({\vec p}\cdot \partial_{\vec q})} e^{t({\nabla\Phi}\cdot \partial_{\vec p})} e^{-{t\over 2}({\vec p}\cdot \partial_{\vec q})} f({\vec q},{\vec p},0).
\end{equation}
This can be evaluated easily
by alternately performing forward and backward FFT's separately in ${\vec q}$ and ${\vec p}$.
Applications such as these argue that FFT packages with built-in multi-dimensional capabilities should allow the user to control the FFT direction in every dimension separately.

\section*{Acknowledgements}
The author thanks Ji Qiang, Alex Friedman, James Amundson, and Alexandru Macridin for helpful discussions.
This work was supported by the Office of Science of the U.S. Department of Energy, Office of High Energy Physics and Office of Advanced Scientific Computing Research,
through the ComPASS project of the SciDAC program, and through funding provided by Fermi National Accelerator Laboratory.
This research used resources of the National Energy Research Scientific Computing Center,
which is supported by the Office of Science of the U.S. Department of Energy under Contract No. DE-AC02-05CH11231.

\appendix

\section{FFT-based Convolutions and Correlations}
Discrete convolutions arise in solving the Poisson equation, as well as in signal processing.
In regard to the Poisson equation, one is typically interested in the following,

\begin{equation}
p_j=\sum_{k=0}^{K-1}r_k x_{j-k}\quad,
\begin{array}{l}
j=0,\ldots,J-1 \\
k=0,\ldots,K-1 \\
j-k=-(K-1),\ldots,J-1 \\
\end{array}
\label{bruteforceconvolution}
\end{equation}
where $x$ corresponds to the free space Green function, $r$ corresponds to the charge density, and $p$ corresponds to the scalar potential.
The sequence $\{p_j\}$ has $J$ elements, $\{r_k\}$ has $K$ elements, and $\{x_m\}$ has $M=J+K-1$ elements.

One can zero-pad the sequences to a length $N\ge M$ and use FFTs to efficiently obtain the $\{p_j\}$ in the unpadded region.
To see this, define a zero-padded charge density, $\rho$,
\begin{equation}
\rho_k=\left\{
\begin{array}{l l}
r_k & \quad \text{if }k=0,\ldots,K-1 \\
0 & \quad \text{if }k=K,\ldots,N-1. \\
\end{array}\right.
\end{equation}
Define a periodic Green function, $\xi_m$, as follows,
\begin{equation}
\xi_m=\left\{
\begin{array}{l l}
x_m & \quad \text{if }m=-(K-1),\ldots,J-1 \\
0 & \quad \text{if }m=J,\ldots,N-K, \\
\xi_{m+iN}=\xi_{m} & \quad \text{for any integer }i 
\end{array}\right.
\label{periodicgreenfunction}
\end{equation}
Now consider the sum
\begin{equation}
{\phi}_j={1\over N}\sum_{k=0}^{N-1} W^{-jk}
                    (\sum_{n=0}^{N-1} \rho_n W^{nk})
                    (\sum_{m=0}^{N-1} \xi_m W^{mk})
~~~~~~0 \le j \le N-1,
\label{fftconvolution}
\end{equation}
where $W=e^{-2\pi i/N}$. This is just the FFT-based convolution of $\{\rho_k\}$ with $\{\xi_m\}$.
Then,
\begin{equation}
{\phi}_j=
          \sum_{n=0}^{K-1}~
          \sum_{m=0}^{N-1} r_n \xi_m
{1\over N}\sum_{k=0}^{N-1} W^{(m+n-j)k}
~~~~~~0 \le j \le N-1.
\end{equation}
Now use the relation
\begin{equation}
\sum_{k=0}^{N-1} W^{(m+n-j)k}= N \delta_{m+n-j,iN}~~~~~(i~\rm an~integer).
\end{equation}
It follows that
\begin{equation}
{\phi}_j=\sum_{n=0}^{K-1}~r_n \xi_{j-n+iN}
~~~~~~0 \le j \le N-1.
\end{equation}
But $\xi$ is periodic with period $N$. Hence,
\begin{equation}
{\phi}_j=\sum_{n=0}^{K-1}~r_n \xi_{j-n}
~~~~~~0 \le j \le N-1.
\label{finaleqn}
\end{equation}
In the physical (unpadded) region, $j\in \left[0,J-1\right]$, so the quantity $j-n$ in Eq.~(\ref{finaleqn}) satisfies $-(K-1)\le j-n \le J-1$.
In other words the values of $\xi_{j-n}$ are identical to $x_{j-n}$. Hence, in the physical region the FFT-based convolution, Eq.~(\ref{fftconvolution}),
matches the convolution in Eq.~(\ref{bruteforceconvolution}).

As stated above, the zero-padded sequences need to
have a length $N \ge M$, where $M$ is the number of elements in the Green function sequence $\left\{x_m\right\}$.
In particular, one can choose $N=M$, in which case the Green function sequence is not padded at all, and only 
the charge density sequence, $\left\{r_k\right\}$, is zero-padded, with $k=0,\ldots,K-1$ corresponding to the physical region
and $k=K,\ldots,M-1$ corresponding to the zero-padded region.


The  above FFT-based approach -- zero-padding the charge density array, and circular-shifting the Green function in accordance with Eq.~(\ref{periodicgreenfunction}) -- will work in general.
In addition, if the Green function is a symmetric function of its arguments,  the value at the end of the Green function array
(at grid point $J-1$)
can be dropped, since it will be recovered implicitly through the symmetry of Eq.~(\ref{periodicgreenfunction}).
In that case the approach is identical to the Hockney method \cite{hockney, eastwoodandbrownrigg,hockneyandeastwood}.

Lastly, note that the above proof that the convolution,  Eq.~(\ref{fftconvolution}), is identical to Eq.~(\ref{bruteforceconvolution}) in the unpadded region, works even when
$W^{-j k}$ and $W^{m k}$  are replaced by $W^{j k}$ and $W^{-m k}$, respectively, in Eq.~(\ref{fftconvolution}). In other words, the FFT-based approach can be used to compute
\begin{equation}
p_j=\sum_{k=0}^{K-1}r_k x_{j+k}\quad,
\begin{array}{l}
j=0,\ldots,J-1 \\
k=0,\ldots,K-1 \\
j-k=-(K-1),\ldots,J-1 \\
\end{array}
\label{bruteforcecorrelation}
\end{equation}
simply by changing the direction of the Fourier transform of the Green function and changing the direction of the final Fourier transform.


\bibliographystyle{model1-num-names}
\bibliography{<your-bib-database>}



\end{document}